\documentclass[preprint,showpacs,preprintnumbers,amsmath,amssymb,nofootinbib]{revtex4}

\usepackage{etex}
\usepackage{amssymb,amsthm,amscd,amsbsy,array}
\usepackage{bm}
\usepackage{soul} 

\usepackage{graphics,graphicx,xcolor}

\usepackage[colorlinks=true, pdfstartview=FitV, linkcolor=blue, citecolor=blue, urlcolor=blue]{hyperref} 
\newcommand{\cyan}{\textcolor{cyan}}
\newcommand{\magenta}{\textcolor{magenta}}
\newcommand{\red}{\textcolor{red}}

\newcommand{\blue}{\textcolor{blue}}

\newcommand{\purple}{\textcolor[rgb]{0.5,0.0,0.5}}
\newcommand{\gb}{\quad\colorbox{green}}

\newcommand{\dgreen}{\textcolor[rgb]{0,0.5,0}}

\newenvironment{redtext}{\color{red}}
{\ignorespacesafterend}
\newenvironment{bluetext}{\color{blue}}{\ignorespacesafterend}

\newenvironment{magentatext}{\color{magenta}}{\ignorespacesafterend}
\newenvironment{cyantext}{\color{cyan}}{\ignorespacesafterend}
\newenvironment{orangetext}{\color{orange}}
{\ignorespacesafterend}

\newcommand{\bmagenta}{\begin{magentatext}}
\newcommand{\emagenta}{\end{magentatext}}
\newcommand{\bcyan}{\begin{cyantext}}
\newcommand{\ecyan}{\end{cyantext}}
\newcommand{\bblue}{\begin{bluetext}}
\newcommand{\eblue}{\end{bluetext}}
\newcommand{\bred}{\begin{redtext}}
\newcommand{\ered}{\end{redtext}}
\newcommand{\borange}{\begin{orangetext}}
\newcommand{\eorange}{\end{orangetext}}

\numberwithin{equation}{section}

\let\ssection=\section
\renewcommand{\section}{\setcounter{equation}{0}\ssection}
\newcommand{\beq}{\begin{equation}}
\newcommand{\eeq}{\end{equation}}

\newcommand{\cA}{{\mathcal{A}}}

\newcommand{\baromega}{\bar{\omega}}
\newcommand{\wpsi}{\widetilde{\psi}}

\def\aand{{\quad\text{and}\quad}}
\def\where{{\quad\text{where}\quad}}

\newcommand{\sign}{\mathrm{sign}}

\def\smallover\#1/\#2{\hbox{$\textstyle\frac{\#1}{\#2}$}} %

\def\JI{{Junker and Inomata\,}}

\def\parag{\hfil\break} 
\def\kikezd{\parag\underbar}

\def\benu{\begin{enumerate}}
\def\eenu{\end{enumerate}}
\def\bitem{\begin{itemize}}
\def\eitem{\end{itemize}}

\def\beq{\begin{equation}}
\def\eeq{\end{equation}}
\def\beqa{\begin{eqnarray}}
\def\eeqa{\end{eqnarray}}
\def\nn{\nonumber}
\def\barray{\left(\begin{array}}
\def\earray{\end{array}\right)}
\def\barraynb{\begin{array}}
\def\earraynb{\end{array}}

\def\IR{{\mathbb{R}}} 

\def\?{{\quad\gb{\fbox{\texttt{?}}\;}}\quad}

\def\p{{\partial}}

\def\v0{\mathbf{0}}

\def\Rarrow{{\quad\Rightarrow\quad}}

\def\p{\partial}

\def \p{{\partial}}

\newcommand{\const}{\mathop{\rm const.}\nolimits}
\newcommand{\half }{\smallover{1}/{2}}

\def\smallover#1/#2{\hbox{$\textstyle\frac{#1}{#2}$}} %
\def\smallcirc{{\raise 0.5pt \hbox{$\scriptstyle\circ$}}}
\def\cabove(#1){\stackrel{\smallcirc}{#1}}
\def\ccabove(#1){\stackrel{\smallcirc\smallcirc}{#1}}
\def\2{{\smallover1/2}}

\def\cA{{\cal A}}
\def\boxit#1{
\vbox{\hrule\hbox{\vrule\kern4pt
\vbox{\kern5pt#1\kern5pt}\kern4pt\vrule}\hrule}
} 

\newcommand{\bigbox}[1]{\fbox{%
\rule[-20pt]{0pt}{45pt}$\;\;\displaystyle{#1}\;\;$}
}
\newcommand{\medbox}[1]{\fbox{%
\rule[-10pt]{0pt}{25pt}$\;\;\displaystyle{#1}\;\;$}%
}

\let\ssection=\section
\renewcommand{\section}
{\setcounter{equation}{0}\ssection}

\def\besub{\begin{subequations}}
\def\esub{\end{subequations}}

\begin{document}

\preprint{\texttt{2105.07374v4 [quant-ph]}}

\title{Time-dependent conformal transformations
\\
and the propagator  for
\\
 quadratic systems
 \footnote{This paper celebrates the 90th birthday of Akira Inomata. 
}
}

\author{
Q.-L. Zhao$^{1}$\footnote{mailto: zhaoqliang@mail2.sysu.edu.cn},
P.-M. Zhang$^{1}$\footnote{mailto:zhangpm5@mail.sysu.edu.cn}
 and 
P. A. Horvathy$^{2}$\footnote{mailto:horvathy@lmpt.univ-tours.fr}\\
}

\affiliation{
${}^1$ School of Physics and Astronomy, Sun Yat-sen University, Zhuhai, (China)
\\
${}^2$  Institut Denis Poisson CNRS/UMR 7013 - Universit\'e de Tours - Universit\'e d'Orl\'eans Parc de Grandmont, 37200, Tours, (France)
}

\date{\today}


\begin{abstract}
The method proposed by Inomata and his collaborators allows us to transform a damped Caldiroli-Kanai oscillator with time-dependent frequency to one with constant frequency and no friction by redefining the time variable, obtained by solving a Ermakov-Milne-Pinney equation. Their mapping ``Eisenhart-Duval'' lifts as a conformal transformation between two appropriate Bargmann spaces. 
The quantum propagator is calculated also by bringing the quadratic system to free form by another time-dependent Bargmann-conformal transformation which generalizes the one introduced before by Niederer and is related to the mapping proposed by Arnold. Our approach allows us to extend  the Maslov phase correction to arbitrary time-dependent frequency. The method is illustrated by the Mathieu profile.
\\[10pt]
\texttt{Symmetry} {\bf 2021} 13, 1866. https://doi.org/10.3390/sym13101866
\end{abstract}

\maketitle

\tableofcontents

\section{Introduction}\label{Intro}

Let us consider a non-relativistic quantum particle with unit mass 
 in $d+1$ space-time dimensions with coordinates $x,t$, given by the natural Lagrangian $L=\half{\dot{x}}^2-V(x,t)$. 
 The wave function is expressed in terms of the propagator, 
\beq
\psi(x'',t'') = \int K(x'',t''|x',t')\psi(x',t')
dx' 
\label{wfprop} 
\eeq 
 which, following Feynman's intuitive proposal \cite{FeynmanHibbs}, is obtained as,
\beq
K(x'',t''|x',t') =\int\exp\big[\frac{i}{\hbar}\cA(\gamma)\big]{\cal D\gamma}\,,
\label{pathintegral}
\eeq
where the (symbolic) integration is over all paths 
$\gamma(t)=\big(x(t),t\big)$ which link the space-time point $(x',t')$ to $(x'',t'')$, and where
\beq
\cA(\gamma)=\int_{t'}^{t''}\!\!\! L\big(\gamma(t),\dot{\gamma}(t),t\big) dt
\label{gclaction}
\eeq
is the {classical action} calculated along  $\gamma(t)$ \cite{FeynmanHibbs,Schulman,KLBbook}. 

The rigourous definition and calculation of \eqref{pathintegral}  is beyond our scope here. 
However the \emph{semiclassical approximation} leads to the van Vleck-Pauli formula \cite{Schulman,KLBbook,semiclassic},
\beq
K(x'',t''|x',t') =\left[\frac{i}{2\pi\hbar}\frac{\p^2\bar{\cA}}{\p x'\p x''}\right]^{1/2} \!\exp\left[\frac{i}{\hbar}\bar{\cA}(x'',t''|x',t')\right]\,,
\label{scprop}
\eeq
where 
$
\bar{\cA}(x'',t''| x',t')\!=\!\displaystyle\int_{t'}^{t''}\!\!\! L(\bar{\gamma}(t),\dot{\bar{\gamma}}(t),t) dt
$
 is the classical action calculated along the (supposedly unique\footnote{This condition is satisfied away from caustics \cite{Schulman,HFeynman,KLBbook}. Morevover \eqref{freeprop}  and \eqref{osciprop} are valid only for $0< T''-T'$ and for 
$0< t''-t'<\pi$, respectively as it will be discussed in sec.\ref{Maslovsec}.}) 
classical path  $\bar{\gamma}(\tau)$ from $(x',t')$ and $(x'',t'')$. This expression involves  data of the classical motion only. We note here also the {van Vleck determinant}  
$\frac{\p^2\bar{\cA}}{\p x'\p x''}\,$ in the prefactor \cite{semiclassic}.

Eqn. \eqref{scprop} is {exact} for a quadratic-in-the-position potentials in $1+1$ dimension 
$ 
V(x,t)=\half\omega^2(t)\,x^2
$ 
that we consider henceforth.

For $\omega\equiv 0$, i.e., for a free non-relativistic particle of unit mass in 1+1 dimensions with coordinates $X$ and $T$, the result is \cite{FeynmanHibbs,Schulman,KLBbook},
\beq
K_{free}(X'',T''|X',T') =\left[\frac{1}{2\pi{i}\hbar(T''-T')}\right]^{1/2} \!\!
\exp\left\{\frac{i}{\hbar}\frac{(X''-X')^2}{2(T''-T')}\right\}\,.
\label{freeprop}
\eeq

An harmonic oscillator with dissipation is in turn described by the Caldirola-Kanai (CK) Lagrangian and equation of motion, respectively  \cite{Caldirola}.
For constant damping and harmonic frequency we have,
\besub
\begin{align}
&L_{CK}= \frac{1}{2}e^{\lambda_0{t}}\Big(\big(\frac{dx}{dt}\big)^2
-\omega_0^2x^2\Big)\,,
\label{CaKaLag}
\\[2pt]
&\frac{d^2x}{dt^2}+\lambda_0\frac{dx}{dt}+\omega_0^2\,x=0\,
\label{CaKaeqmot}
\end{align}
\label{CaKa}
\esub
with $\lambda_0=\const > 0$ and $\omega_0=\const$.
A lengthy calculation then yields the exact propagator \cite{Schulman,KLBbook,Dekker,Khandekar,Um02} 
\besub
\begin{align}
&K_{CK}(x'',t''|x',t')=
\left[\frac{\Omega_0\,e^{\frac{\lambda_0}{2}(t''+t')}}
{2\pi i\hbar\,\sin\big[\Omega_0(t''-t')\big]}\right]^{\frac{1}{2}}\quad \times
\label{osciprop} 
\\[6pt]
&\qquad\exp\left\{\frac{i{\Omega_0}}{2\hbar\,\sin\big[{\Omega_0}(t''-t')\big]}
\left[(x''^2e^{\lambda_0t''}
+x'^2e^{\lambda_0t'}
)\cos\big[{\Omega_0}(t''-t')\big]
-2x''x'e^{\lambda_0\frac{t''+t'}{2}}
\right]\right\}, 
\nonumber 
\\[10pt]
&{\Omega}_0^2=\omega_0^2-\smallover{1}/{4}\lambda_0^2\,,
\end{align}
\label{Oomega}
\esub
where an irrelevant phase factor was dropped. 

\goodbreak
Inomata and his collaborators \cite{JunkerInomata, Cai82, Cai} generalized \eqref{Oomega} to time-dependent frequency by redefining time, $t \to \tau$, which allowed them to transform the time-dependent problem to one with constant frequency (see sec.\ref{JIsec}).
Then they follow by what they call a  ``\emph{time-dependent conformal transformation}'' $(x, t)\to (X, T)$ such that
\beq
x = f(T)\,X(T)\exp\left[\half\lambda_0T\right],
\;\;\;
t = \,g(T)\,,
\where
f^2(T) =\,\frac{dg}{dT}\,,
\label{Inomatatransf}
\eeq
which allows them to derive the propagator from the free expression \eqref{freeprop}. When spelled out, \eqref{Inomatatransf} boils down to a generalized version, \eqref{XTtrans}, of the correspondence found by  Niederer \cite{Niederer73}.

It is legitimate to wonder~: \textit{in what sense are these transformations ``conformal''}~? In sec.\ref{Bargmannsec} we explain that in fact \emph{both} mappings can  be interpreted in the Eisenhart-Duval (E-D) framework as  conformal transformations between two appropriate Bargmann spaces
\cite{Eisenhart,DBKP,BurdetOsci,DGH91,dissip}. 
Moreover, the change of variables $x,t \to X, T$ is  a special case of the one put forward by Arnold \cite{Arnold}, and will be shown to be  convenient to study explicitly time-dependent systems. 

A  bonus is  the extension to arbitrary time-dependent frequency $\omega(t)$ of the Maslov phase correction \cite{Maslov,Arnold67,SMaslov,Burdet78,Schulman,semiclassic,HFeynman,BurdetOsci,RezendeJMP} even when no explicit solutions are available (see sec.\ref{Maslovsec}). 

In sec.\ref{Mathieusec} we illustrate our theory by the time-dependent Mathieu profile $\omega^2(t)= a-2q\cos 2t\,,\, a, b \const$ whose direct analytic treatment is complicated.

\section{The Junker-Inomata derivation of the propagator}\label{JIsec}

Starting with a general quadratic Lagrangian in 1+1 spacetime dimensions with coordinates $\tilde{x}$ and $t$, \JI derive the equation of motion \cite{JunkerInomata}
\beq
\ddot{\tilde{x}}+\dot{\lambda}(t)\dot{\tilde{x}}+\omega^2(t)\tilde{x}=F(t)\,, 
\label{eqq}
\eeq
which  describes a non-relativistic particle of unit mass with  dissipation $\lambda(t)$.
The  driving force $F(t)$ can be eliminated by subtracting a particular solution $h(t)$ of \eqref{eqq}, 
$ 
x(t)=\tilde{x}(t)-h(t), 
$ 
in terms of which \eqref{eqq} becomes homogeneous,
\begin{eqnarray}
\ddot{x}+\dot{\lambda}(t)\dot{x}+\omega^2(t)x=0\,. 
\label{eqx}
\end{eqnarray}
This equation can be obtained  from the time-dependent generalization of  \eqref{CaKaLag},
\begin{eqnarray}
L_{CK}=\frac{1}{2}e^{\lambda(t)}[\dot{x}^2-\omega^2(t)x^2]\,. 
\label{CKLag}
\end{eqnarray}

The friction can be eliminated by setting $x(t) = y(t)\,e^{-\lambda(t)/2}$ which yields an harmonic oscillator with no friction but with shifted frequency \cite{Aldaya}, 
\beq
\ddot{y}+\Omega^2(t)y=0
\;\where\;
\Omega^2(t)=\omega^2(t)-\frac{\dot{\lambda}^2(t)}{4}-\frac{\ddot{\lambda}(t)}{2}\,. 
\label{SLeq}
\eeq
For $\lambda(t)=\lambda_0t$ and $\omega=\omega_0=\const$, for example, we get a usual harmonic oscillator with constant shifted frequency, 
$ 
\Omega^2=\omega_0^2-{\lambda_0^2}/{4} =\const
$ 

 The frequency is in general time-dependent, though, $\Omega=\Omega(t)$, therefore \eqref{SLeq} is a \emph{Sturm-Liouville equation} that can be solved analytically only in exceptional cases. 

\JI \cite{JunkerInomata} follow  another, more subtle path.
Eqn. \eqref{eqx} is a linear equation with time-dependent coefficients whose solution can be searched for within the Ansatz
\footnote{A similar transcription was proposed, independently, also by Rezende \cite{RezendeJMP}.}
\beq
\medbox{
x(t)=\rho(t)\Big(Ae^{i{\bar{\omega}}\tau(t)}+Be^{-i{\bar{\omega}}\tau(t)}\Big) \,,
}
\label{xAnsatz}
\eeq
where $A$, $B$ \underline{and ${\bar{\omega}}$} are constants and $\rho(t)$ and $\tau(t)$ functions to be found.
Inserting \eqref{xAnsatz} into \eqref{eqx}, putting the coefficients of the exponentials to zero, separating real and imaginary parts and absorbing a new integration constant into $A,\,B$ provides us with the coupled system for 
$\rho(t)$ and $\tau(t)$,
\besub
\begin{align}
&\ddot{\rho}+\dot{\lambda}\dot{\rho}+(\omega^2(t)-{\bar{\omega}}^2\dot{\tau}^2)\rho=0,
\label{eqrho} 
\\[2pt]
&\dot{\tau}(t)\,\rho^2(t) \,e^{\lambda(t)}=1\,.
\label{trlconstr}
\end{align}
\label{rhotaueq}
\esub
Manifestly $\dot{\tau}>0$.
Inserting ${\dot{\tau}}$ into \eqref{eqrho} then yields the \textit{Ermakov-Milne-Pinney} (EMP) equation \cite{EMP}
 with time-dependent coefficients,
\beq
\ddot{\rho}+\dot{\lambda}\dot{\rho}+\omega^2(t)\rho=
\frac{e^{-2\lambda(t)}\baromega^2}{\rho^3}\,.
\label{genErmakov}
\eeq
 We note for later use that eliminating $\rho$ would yield instead
\beq
\bar{\omega}^2=\frac{1}{\dot{\tau}^2}\left(
\omega^2(t)
-\frac{1}{2}\frac{\dddot{\tau}}{\dot{\tau}}
+\frac{3}{4}\left(\frac{\ddot{\tau}}{\dot{\tau}}\right)^2
-\frac{\ddot{\lambda}}{2}-\frac{\dot{\lambda}^2}{4}\right)\,.
\label{labjegy}
\eeq
Conversely, the constancy of the r.h.s. here can be verified using the eqns \eqref{rhotaueq}. Equivalently, starting with the Junker-Inomata condition \eqref{Inomatatransf},
\beq
\omega^2(t)=\frac{\ddot{f}}{f}-2\frac{\dot{f}^2}{f^2}
+
\frac{\dot{\lambda}^2}{4}+\frac{\ddot{\lambda}}{2}\,.  
\label{fomega}
\eeq
\goodbreak

To sum up, the strategy we follow is \cite{JunkerInomata,Galajinsky}~:
\benu
\item
to solve first the EMP equation \eqref{genErmakov} for $\rho$,

\item
to integrate \eqref{trlconstr},
\beq
\tau(t) = \int^t\!\frac{e^{-\lambda(u)}}{\rho^2(u)}\,du\,.
\label{tauint}
\eeq
\eenu
Then the trajectory is given by \eqref{xAnsatz}.

\goodbreak
Junker and Inomata show, moreover, that substituting into \eqref{CKLag} the new coordinates
\beq
\medbox{
T=\frac{\tan{[{\bar{\omega}}\,\tau(t)]}}{{\bar{\omega}}}\,,
\qquad
X=x\,e^{\frac{\lambda(t)}{2}}\dot{\tau}(t)^{\frac{1}{2}}\sec{[{\bar{\omega}}\,\tau(t)]},  
} 
\label{XTtrans}
\eeq
allows us to present the Caldirola-Kanai action  as \footnote{Surface terms do not change the classical equations of motion and multiply the propagator by an unobservable phase factor, and will therefore dropped.},
\beq
\cA_{CK}= \displaystyle\int_{t'}^{t''}\!\! L_{CK}dt =\int_{T'}^{T''}\!\frac{1}{2}
\big(\frac{dX}{dT}\big)^2dT\,,
\label{freeaction}
\eeq
where we recognize the \emph{action of a free particle} of unit mass. One checks also directly that $X,T$ satisfy the free equation as they should. 
The conditions \eqref{Inomatatransf} are readily verified. 

The coordinates $X$ and $T$ describe a free particle, therefore  the propagator is \eqref{freeprop}  (as anticipated by our notation). The clue of \JI  \cite{JunkerInomata} is that, conversely, trading $X$ and $T$ in \eqref{freeprop} for $x$ and $t$ allows to derive the propagator for the CK oscillator (see also \cite{Um02}, sec.5.1),
\footnote{The extension of \eqref{JIprop} from $0 < \bar{\omega}(\tau''-\tau') < \pi$  to all $t$  \cite{Schulman,HFeynman,Um02,KLBbook},  
will be discussed in sec.\ref{Maslovsec}.},
\beqa
&&K_{osc}(x'',t''|x',t') = 
\left[\frac{{\bar{\omega}}e^{\frac{\lambda''+\lambda'}{2}}(\dot{\tau}''\dot{\tau}')^{\frac{1}{2}}}
{2\pi i\hbar\,\sin[{\bar{\omega}}(\tau''-\tau')]}\right]^{\frac{1}{2}} \;\times 
\label{JIprop}
\\[6pt]
&&\quad\exp\left\{ \frac{i{\bar{\omega}}}{2\hbar\,\sin[{\bar{\omega}}(\tau''-\tau')]}\left[(x''^2e^{\lambda''}\dot{\tau}''+x'^2e^{\lambda'}\dot{\tau}')\cos[{\bar{\omega}}(\tau''-\tau')]
-2x''x'e^{\frac{\lambda''+\lambda'}{2}}(\dot{\tau}''\dot{\tau}')^{\frac{1}{2}}\right]\right\}, \quad
\nonumber 
\eeqa
where we used the shorthands $\lambda''=\lambda(t''),\,  \tau''=\tau(t'')$, etc. 

This remarkable formula says that in terms of ``redefined time'', $\tau$, the problem is essentially one with constant frequency.
Eqn. \eqref{JIprop} is still implicit, though, as it requires to solve first the coupled system \eqref{rhotaueq} that we can do only in particular cases. 
\begin{itemize}
\item
When $\lambda(t)=\lambda_0t$ where
$\lambda_0=\const\geq0 $ eqn \eqref{eqx} describes a time-dependent  oscillator with constant friction,
\begin{eqnarray}
\ddot{x}+\lambda_{0}\,\dot{x}+\omega^2(t)x=0. 
\label{tdo1}
\end{eqnarray}
Then setting
$R(t)=\rho(t)\,e^{\lambda_0 t/2}$  eqns \eqref{rhotaueq} 
provide us with the EMP equation for $R$, cf. \eqref{genErmakov},
\beq
\ddot{R}+\Omega^2(t)R-\frac{\baromega^2}{R^3}=0,
\where
\Omega^2(t)=\omega^2(t)-\frac{\lambda_0^2}{4}\,.
\label{EMPeq}
\eeq
\item
If, in addition, the \emph{frequency} is \emph{constant}
$ 
\omega(t)=\omega_0=\const,
$ 
then eqn \eqref{EMPeq} is solved algebraically by 
\beq
\baromega^2=\omega_{0}^2
-\lambda_{0}^2/4,
\quad
R=1
\Rarrow \rho(t)=e^{-\lambda_0t/2},  
\quad
\tau(t)=t. 
\label{omegaomega}
\eeq
 Thus $x(t)$ is a linear combination of 
$
e^{-\half\lambda_0t}\sin\baromega t$ and $e^{-\half\lambda_0t}\cos\baromega t.
$
The space-time coordinate transformation of $(x,t) \to (X,T)$ in \eqref{XTtrans} simplifies to the friction-generalized form of that of Niederer \cite{Niederer73},
\begin{eqnarray}
T=\frac{\tan(\baromega {t})}{\baromega}\,,
\quad 
X=x\exp\left(\frac{1}{2}\lambda_{0}t\right)\sec(\baromega{t}),  
\label{Ntransf} 
\end{eqnarray}
for which the general expression \eqref{JIprop} reduces to \eqref{Oomega} when $\lambda_0=0$.

\item
When the oscillator is turned off, $\omega_0=0$ but $\lambda_0>0$, we have motion in a dissipative medium. The coordinate transformation propagator \eqref{XTtrans} and \eqref{JIprop} become 
\beq
X=\frac{2x}{1+\exp\left(-\lambda_{0}t\right)}\,,
\qquad
T=
\frac{2}{\lambda_{0}}\frac{1-\exp(-\lambda_{0}t)}{1+\exp(-\lambda_{0}t)}\, 
\label{ystrans1}
\eeq
and 
\beq
\begin{array}{lll}
K_{diss}(x'',t''| x',t')&=&
\left[\displaystyle\frac{\lambda_{0}}
{2\pi i\hbar[\exp(-\lambda_{0}t')-\exp(-\lambda_{0}t'')]} \right]^{\frac{1}{2}}  
\\[14pt]
&\quad 
&\times\exp\left\{
\displaystyle\frac{i\lambda_{0}}{2\hbar}\frac{(x''-x')^2}{\exp(-\lambda_{0}t')-\exp(-\lambda_{0}t'')} \right\}\,,
\end{array} 
\label{npro}
\eeq
respectively.
A driving force $F_0$ (e.g. terrestrial gravitation) could be added and then removed by  $x\to x+({F_{0}}/{\lambda_{0}})t$.
\end{itemize}
Further examples can be found in \cite{Cai82,Cai}. An explicitly time-dependent example will be presented in sec. \ref{Mathieusec}.

\section{The Eisenhart-Duval lift}\label{Bargmannsec}

Further insight can be gained by ``Eisenhart-Duval (E-D) lifting'' the system to one higher dimension to what is called a ``Bargmann space'' \cite{Eisenhart,DBKP,BurdetOsci,DGH91,dissip}.
 The latter is a $d+1+1$ dimensional manifold endowed with a Lorentz metric whose general form is
\begin{equation}
g_{\mu\nu}dx^{\mu}dx^{\nu}=g_{ij}(x,t)dx^{i}dx^{j}+2dtds-2 V(x,t)dt^{2}\,,
\label{genBmetric}
\end{equation}
which carries a covariantly constant null Killing vector $\p_s$. Then~: 

\vskip-2mm
\kikezd{Theorem 1 \cite{DBKP,DGH91}}~: \textit{Factoring out the foliation generated by $\p_s$ yields  a non-relativistic space-time in $d+1$ dimensions. 
 Moreover, the null geodesics of the Bargmann metric $g_{\mu\nu}$ 
 project to ordinary space-time consistently with Newton's equations. Conversely, if $(\gamma(t),t)$ is a solution of the non-relativistic equations of motion, then its null lifts to Bargmann space are\vspace{-1mm}
\beq
\big(\gamma(t),t,s(t)\big),
\quad
s(t)=s_0-{\cal A}(\gamma)=s_0-\int^t\!\!L(\gamma(r),r)dr
\label{Blift}
\eeq \vspace{-1mm}
where $s_0$ is an arbitrary initial value.
}
\goodbreak
 
\smallskip
Let us consider, for example, a particle of unit mass with the Lagrangian of  
\beq
L= \displaystyle\frac{1}{2\alpha(t)} g_{ij}(x^k) \dot x^i \dot x^j - \beta(t) V(x^i,t), 
\label{abLag}
\eeq
where  $g_{ij}(x^k)dx^idx^j$ is a  positive metric on a curved configuration space $Q$ with local coordinates $x^i$, $i=1,\ldots,d$. 
 The coefficients  $\alpha(t)$ and $\beta(t)$ may depend on time $t$ and $V(x^i,t)$ is some (possibly time-dependent)  scalar potential. The associated equations of motion are
\beq
\frac{d^2x^i}{dt^2}+\Gamma^i_{jk}\frac{dx^j}{dt}\frac{dx^k}{dt} - \frac{\dot\alpha}{\alpha} \frac{dx^i}{dt}
= 
-\alpha \beta g^{ij}\p_ j V\,,
\label{alphabetaeq}
\eeq
where 
the $\Gamma^i_{jk}$ are the Christoffel symbols of 
the metric $g_{ij}$. 
 For  $d=1$, $g_{ij}=\delta_{ij}$ and $V= \half \omega^2(t) x^2$ for $\alpha = \beta = 1$ resp. for 
$\alpha = \beta^{-1} = e^{-\lambda(t)}$
we get a (possible time-dependent) 1d oscillator without resp. with friction, eqn. \eqref{CaKa}, \cite{Caldirola,Dekker,Aldaya}.  

Equation \eqref{alphabetaeq} can also be obtained by projecting a null-geodesic of $d+1+1$ dimensional Bargmann spacetime with coordinates 
$(x^\mu)=({x}^i,t,s) $, whose metric is
\begin{equation}
g_{\mu\nu}dx^{\mu}dx^{\nu}=\frac{1}{\alpha}g_{ij}dx^{i}dx^{j}+2dtds-2\beta Vdt^{2}.
\label{alphabetametric}
\end{equation}%
For $\alpha = \beta^{-1} = e^{-\lambda(t)}$ we recover  \eqref{eqx}. 

Choosing $\lambda(t)= \ln m(t)$ would describe motion with a time-dependent mass $m(t)$. The friction can be removed by the conformal rescaling $x \to y = \sqrt{m}\,x$ and the null geodesics of the rescaled metric describe, consistently with \eqref{SLeq}, an oscillator with no friction but with time-dependent frequency, 
$\Omega^2=\omega^2- \frac{\ddot{m}}{2m}+\big(\frac{\dot{m}}{2m})^2$
\cite{Cheng85}. 

The friction term $-(\dot\alpha/\alpha)\dot{x}^i$ in \eqref{alphabetaeq}  can be removed also by introducing a new time-parameter $\tilde{t}$, defined by
$d\tilde{t}=\alpha\, dt$ \cite{dissip}.
 For  $\lambda(t)=\lambda_0t$, for example, putting ${\tilde{t}}=-e^{-\lambda_0t}/\lambda_0$ and  eliminates the  friction -- but it does it  at the price of getting manifestly time-dependent frequency \cite{Ilderton,Conf4GW}
\beq
\frac{d^2x}{d\tilde{t}^2} +\tilde{\Omega}^2(\tilde{t})x = 0 \,,\qquad 
\tilde{\Omega}^2(\tilde{t}) = 
\frac{\omega^2}{\tilde{t}^2\lambda_0^2}\,. 
\label{invtsqfreq}
\eeq
\subsection{The Junker-Inomata Ansatz as a conformal transformation}\label{BJIsec}

The approach outlined in sec.\ref{JIsec} admits a Bargmannian interpretation. For simplicity we only consider the frictionless case $\lambda=0$.

\kikezd{Theorem 2}~: \textit{The Junker-Inomata method of converting the time-dependent system into one with constant frequency by switching from ``real'' to ``fake time'', 
\beq
t \to \tau(t), \qquad \xi = \sqrt{\dot{\tau}}\,{x}\,
\label{txtauxi}
\eeq 
induces a conformal transformation  between the Bargmann metrics
\vspace{-3mm}
\besub
\begin{align}
&dx^2+2dtds-\omega^2(t)x^2dt^{2}
&\text{frequency}\qquad&\omega^2(t)
\label{omegatmetric}
\\
&d\xi^2 +2d\tau d\sigma-\bar\omega^2\xi^2d\tau^{2}\,,
&\text{frequency}\qquad &\bar\omega=\const
\label{taumetric}
\end{align}
\label{2Bmetrics}
\esub\vspace{-4mm}
\beq
\bigbox{
 d\xi^2+2d\tau d\sigma-\bar\omega^2\xi^2d\tau^{2}=
 \dot{\tau}(t)\Big(dx^2+2dtds-\omega^2(t)x^2dt^{2}
 \Big)\,.
 }
\label{otoconst}
\eeq
}

\vskip1mm
\noindent{\textit{Proof}}~: Putting $\mu=\ln\dot{\tau}$ allows us to present the constant-frequency $\baromega$ \eqref{labjegy} as
\beq
\baromega^2=\dot\tau^{-2}\big(\omega^2(t)-\half\ddot{\mu}+
\smallover{1}/{4}\dot{\mu}^2\big).
\label{omegabarmu}
\eeq  

Then with the notation $\stackrel{\smallcirc}{\xi}={d\xi}/{d\tau}$ we find,
\begin{eqnarray}
\stackrel{\smallcirc}{\xi}{\strut}^2=\dot{\tau}^{-1}\left[\dot{x}^2+\frac{1}{4}\dot{\mu}^2x^2-\frac{1}{2}\ddot{\mu}x^2+\frac{d}{dt}\left(\frac{1}{2}\dot{\mu}x^2\right)\right]\,.
\nn
\end{eqnarray}

Let us now recall that the null lift to Bargmann space of a space-time curve is obtained by subtracting the classical action as vertical coordinate,
\begin{eqnarray}
d\sigma=-L(\xi,\stackrel{\smallcirc}{\xi},\tau)d\tau=
-\half\big(\stackrel{\smallcirc}{\xi}{\strut}^2-\bar{\omega}^2\xi^2\big)d\tau\,.
\end{eqnarray}
Setting here $\xi=\strut{}\dot{\tau}^{1/2}\,x$ and dropping surface terms yields, using the same procedure
for the time-dependent-frequency case, 
\beq
d\sigma = ds = -\half\big(\dot{x}^2-\omega^2(t)x^2\big)dt
\label{dsds}
\eeq
up to surface terms.
Then inserting all our formulae into \eqref{omegatmetric} and \eqref{taumetric} yields \eqref{otoconst}, as stated.
In Junker-Inomata language \eqref{Inomatatransf},  $f(t)=\dot{\tau}^{1/2}\sec(\baromega\tau),\, g(t)=(\baromega)^{-1}\tan (\baromega\tau)$\,.

\goodbreak
Our investigation have so far concerned classical aspects. Now we consider what happens quantum mechanically. Restricting our attention at $d=1$ space dimensions as before \footnote{In $d > 2$ conformal-invariance requires to add  a scalar curvature term to the Laplacian.} we posit that the E-D lift $\wpsi$ of a wave function $\psi$ be equivariant,
\beq
\wpsi(x,t,s) = e^{\frac{i}{\hbar}s}\psi(x,t)
\Rarrow
\p_s\wpsi=\frac{i}{\hbar}\wpsi\,.
\label{liftedwf}
\eeq
Then the massless Klein-Gordon equation for $\wpsi$ associated with the $1+1+1=3$ d Barmann- metric  implies the Schr\"odinger equation in 1+1 d,
\beq
\Delta_g \,\wpsi = 0 \Rarrow i\p_t\psi = \big[-\frac{\hbar^2}{2} \Delta_x +V(x,t)\big]\psi\,
\eeq
where $\Delta_g$ is the Laplace-Beltrami operator associated with the metric. In $d=1$ it is of course $\Delta_x = \p_x^2$ .

A conformal diffeomorphism $(X,T,S) \to \tilde{f}(X,T,S)=(x,t,s)$
with conformal factor $\sigma^2_f$,
$
\tilde{f}^*g_{\mu\nu} = \sigma^2_f{\,}g_{\mu\nu},
$
projects to a space-time transformation $(X,T) \to {f}(X,T)=(x,t)$. It is implemented on a wave function lifted to Bargmann space as
\beq
\wpsi(x,t,s) = \sigma_f^{-1/2}\wpsi(X,T,S)
\label{implement}
\eeq
In secs.\ref{Npropsec} these formulae will be applied to the Niederer map \eqref{Nkdef}.

\goodbreak
\subsection{The Arnold map}\label{Arnoldsec}

The general damped harmonic oscillator with time-dependent driving force $F(t)$ in 1+1 dimensions, \eqref{eqq},
\begin{equation}
\ddot{x}+\dot{\lambda}\dot{x}+\omega^{2}\left(t\right) x=F(t)\,,
\label{gendempx}
\end{equation}%
can be solved by an \emph{Arnold transformation} \cite{Arnold} which ``straightens the trajectories'' \cite{Aldaya,LopezRuiz,dissip}. To this end one introduces new coordinates,%
\begin{equation}
T=\frac{u_{1}}{u_{2}}\,,
\quad 
X=\frac{x-u_{p}}{u_{2}}\,,
\label{ArnoldXT}
\end{equation}%
where $u_{1}$ and $u_{2}$ are solutions of the associated homogeneous equation \eqref{gendempx} with $F\equiv0$ %
 and $u_{p}$ is a particular solution of the full equation
\eqref{gendempx}. %
It is worth noting that \eqref{ArnoldXT} allows to check, independently, the Junker-Inomata criterion in \eqref{Inomatatransf}.
The initial conditions are chosen as,%
\begin{equation}
u_{1}\left( t_{0}\right) =\dot{u}_{2}\left( t_{0}\right) =0,\ \ \dot{u}%
_{1}\left( t_{0}\right) =u_{2}\left( t_{0}\right) =1,\ \ u_{p}\left(
t_{0}\right) =\dot{u}_{p}\left( t_{0}\right) =0.
\label{initcond}
\end{equation}%
Then in the new coordinates the motion becomes free \cite{Arnold}, %
\begin{equation}
X(T) = a T+b\,,
\qquad a, b =\const
\label{aTb}
\end{equation}

Eqn \eqref{gendempx} can be obtained by projecting a null geodesic of the Bargmann metric 
\begin{equation}
{g}_{\mu\nu}d{x}^{\mu}d{x}^{\nu}=e^{\lambda(t)}dx^{2}+2dtds-2e^{\lambda(t)}\left(\frac{1}{2}\omega(t)
^{2}x^{2}-F(t)x\right)dt^{2}\,.
\label{gendempB}
\end{equation}
Completing  \eqref{ArnoldXT} by %
\begin{equation}
S =s+e^{\lambda }u_{2}^{-1}\left( \frac{1}{2}\dot{u}_{2}x^{2}+\dot{u}%
_{p}x\right) +g\left( t\right)
\where
\dot{g}=\frac{1}{2}e^{\lambda}\left(\dot{u}_{p}^{2}-\omega
^{2}u_{p}^{2}+2Fu_{p}\right)\,
\label{Sfroms}
\end{equation}%
 lifts the Arnold map to  Bargmann spaces, $(x,t,s) \to (X,T,S)$ 
 \footnote{In the Junker-Inomata setting \eqref{Inomatatransf},  $f=u_2e^{-\lambda/2}$ and $ g(t)=u_{1}/u_{2}$.}, 
\begin{equation}
{g}_{\mu\nu}d{x}^{\mu}d{x}^{\nu}=e^{\lambda(t)}u_{2}^{2}(t)\big(dX^{2}+2dTdS \big) \,.
\label{BArnold}
\end{equation}
The oscillator metric \eqref{gendempB} is thus carried conformally to the free one, generalizing earlier results \cite{DBKP,BurdetOsci,DHP2}.
 For the damped harmonic oscillator with 
$\lambda(t)=\lambda_0t$ and  $F(t) \equiv 0$, $u_p\equiv 0$ is a particular solution.
 When $\omega=\omega_0=\const$, for example, 
\beq
u_1=e^{-{\lambda_0}t/2}\frac{\sin\Omega_0 t}{\Omega_0},
\; 
u_2=e^{-{\lambda_0}t/2}\big(\cos\Omega_0 t+\frac{\lambda_0}{2\Omega_0}\sin\Omega_0 t\big)\,,
\quad
\Omega_0^2=\omega_0^2-{\lambda_0^2}/{4}\;
\eeq
are two independent solutions of the homogeneous equation 
with initial conditions \eqref{initcond} and provide us with 
\begin{subequations}
\begin{align}
T&=\frac{\sin\Omega_0 t}{\Omega_0(\cos\Omega_0 t+\frac{\lambda_0}{2\Omega_0}\sin\Omega_0 t)},
\\[4pt]
X&=\frac{e^{{\lambda_0}t/2} \, x}{\cos\Omega_0 t+\frac{\lambda_0}{2\Omega_0}\sin\Omega_0 t},
\\[4pt]
S&=s-{\half}e^{{\lambda_0}t} x^2\,\big(\frac{\omega_0^2}{\Omega_0}\big)\,\frac{\sin\Omega_0 t}{\cos\Omega_0 t+\frac{\lambda_0}{2\Omega_0}\sin\Omega_0 t}\,.
\end{align}
\label{Arnoldtransf}
\end{subequations}
In the undamped case $\lambda_0=0$ thus $\Omega_0=\omega_0$, and (\ref{Arnoldtransf}) 
 reduces to that of Niederer \cite{Niederer73} lifted to Bargmann space \cite{BurdetOsci,DGH91},
\beq
T=\frac{\tan\omega_0 t}{\omega_0},
\qquad
X=\frac{x}{\cos\omega_0 t},
\qquad
S=s-{\half}x^2\omega_0 \tan\omega_0 t.
\label{NiedEisen}
\eeq

The Junker-Inomata construction in sec.\ref{JIsec} can be viewed as a particular case of the Arnold transformation.   
We choose $u_p\equiv0$ and the two independent solutions%
\begin{equation}
u_{1}=e^{-\lambda/2}\,\dot{\tau}^{-1/2}\,\frac{\sin \bar{\omega}\tau}{\bar{\omega}},
\qquad
 u_{2}=e^{-\lambda/2} 
\,\dot{\tau}^{-1/2} \,\cos \bar{\omega}\tau\,.
\label{u1u2rt}
\end{equation}%
The initial conditions \eqref{initcond} at $t_0=0$ imply
$\tau(0)= \dot{\rho}(0) =0,\, \rho(0) =\dot{\tau}(0) =1.$
Then spelling out \eqref{Sfroms}, 
\begin{equation}
S =s-\frac{1}{2}e^{\lambda}\left(\bar{\omega}\dot{\tau}\tan 
\bar{\omega}\tau +\frac{1}{2}\dot{\lambda}+\frac{1}{2}\frac{\ddot{\tau}}{\dot{\tau}}\right) x^{2}
\label{SJI}
\end{equation}%
completes the lift of \eqref{XTtrans} to Bargmann spaces. 
In conclusion, the one-dimensional damped harmonic oscillator is described by the conformally flat Bargmann metric, %
\beq
\medbox{
{g}_{\mu\nu}d{x}^{\mu}d{x}^{\nu} 
=\frac{\cos^{2}\bar{\omega}\tau }{\dot{\tau}}\big(dX^{2}+2dTdS\big).
}
\label{Arnoldconfflat}
\eeq

The metric \eqref{Arnoldconfflat} is manifestly conformally flat, therefore its geodesics are those of the free metric, $X(T)= aT+b$.
 Then using \eqref{ArnoldXT} with \eqref{u1u2rt} yields 
\begin{equation}
\medbox{
x(t)=
e^{-\lambda(t)/2} \,\dot{\tau}^{-1/2}(t)\left(a\frac{
 \sin[\bar{\omega}\tau(t)]}{\bar\omega} +b\,\cos[\bar{\omega}\tau(t)]\, \right) .
}
\label{xtaueq}
\end{equation}
The bracketed quantity here describes a constant-frequency oscillator with ``time'' 
$\tau(t)$. The original position, $x$, gets a time-dependent ``conformal'' scale factor.

\section{The Maslov correction}\label{Maslovsec}

As mentioned before, the semiclassical formula \eqref{Oomega} is correct only in the first oscillator half-period, $0<t''-t' < \pi/\Omega_0$.   
Its extension for all $t$  involves the \emph{Maslov correction}.        
In the constant-frequency case with no friction, for example, assuming that 
 $\Omega_0(t''-t'')/\pi$ is {not} an integer,
  we have \cite{Schulman,KLBbook,HFeynman},
\beqa
K^{ext}(x'',t''|x',t')&=&
\left[\frac{\Omega_0}
{2\pi \hbar\,\big|\sin \Omega_0(t''-t')\big|}\right]^{\frac{1}{2}}\times \,e^{-i\frac{\pi}{4}(1+2\ell)}
\label{constMprop}
\\[6pt]
&&\exp\left\{
\frac{i\Omega_0}{2\hbar\,\sin \Omega_0(t''-t')}
\left[
({x''}^2+{x'}^2)\cos \Omega_0(t''-t')-2x''x'\right]
\right\}\,,\quad
\nn
\eeqa
where the integer 
\beq
\ell = {\rm Ent} \big[\frac{\Omega_0(t''-t')}{\pi}\big]\,
\label{Maslovindex}
\eeq
 is called as the \emph{Maslov index} (where 
 ${\rm Ent}[x]$ is the integer part of $x$). $\ell$ counts the completed half-periods, and is related also to the \emph{Morse index} which counts the negative modes of $\p^2{\cal A}/\p x'\p x''$ \cite{semiclassic}.  

Now we generalize \eqref{constMprop}  to \emph{time-dependent} frequency~:

\kikezd{Theorem 4}~: \textit{In terms of $\baromega$ and $\tau$  introduced in sec.\ref{JIsec},}
\begin{itemize}
\item
\textit{Outside caustics, i.e., for $\baromega(\tau''-\tau') \neq \pi  \ell$, the propagator for the harmonic oscillator with time-dependent frequency and friction is 
\begin{eqnarray}
&&\hskip-5mm
K^{ext}(x'',t''|x',t')=\left[\frac{\bar{\omega}e^{\frac{\lambda''+\lambda'}{2}}(\dot{\tau}''\dot{\tau}')^{\frac{1}{2}}}{2\pi\hbar|\sin\bar{\omega}(\tau''-\tau')|}\right]^{1/2}
\exp\left\{-\frac{i\pi}{2}\left(\frac{1}{2}+{\rm Ent} \big[\frac{\bar{\omega}(\tau''-\tau')}{\pi}\big] \right) \right\}  
\label{JIMaslov} 
\\[5pt]
&&\times\exp\left\{\frac{i\bar{\omega}}{2\hbar\sin\bar{\omega}(\tau''-\tau')} [(x''^2e^{\lambda''}\dot{\tau}''+x'^2e^{\lambda'}\dot{\tau}')\cos[\bar{\omega}(\tau''-\tau')]-2x''x'e^{\frac{\lambda''+\lambda'}{2}}(\dot{\tau}''\dot{\tau}')^{\frac{1}{2}}]\right\}\,\quad\; 
\nonumber 
\end{eqnarray}
}
\item
\textit{
At caustics, i.e., for 
\beq
\medbox{
\baromega(\tau''-\tau')=\pi\,\ell,\quad \ell = 0, \pm1,\dots
\,}
\label{Deltataupiell}
\eeq
 we have instead} \cite{HFeynman,KLBbook},
\begin{eqnarray}
&&K^{ext}\Big(x'',x',|\tau''-\tau'=
\frac{\pi}{\bar{\omega}}\ell\Big) =
\big[e^{\frac{\lambda''+\lambda'}{2}}(\dot{\tau}''\dot{\tau}')^{\frac{1}{2}}\big]^{1/2} 
\label{JIcaustic}
\\[4pt]
&&\qquad \times \exp\left(-\frac{i\pi\ell}{2}\right)\delta\Big(x'\exp(\lambda'/2)\dot{\tau}'^{1/2}-(-1)^k x''\exp(\lambda''/2)\dot{\tau}''^{1/2}\Big)\,.
 \nonumber
\end{eqnarray}
\end{itemize}

\smallskip
\noindent{\textit{Proof}}~: 
In terms of the redefined coordinates
\beq
\tau=\tau(t)
\aand
\xi=x\exp\left[\frac{\lambda(t)}{2}\right]\dot{\tau}^{1/2}(t),
\label{Zhao7}
\eeq
cf. \eqref{txtauxi}  and using the notation $\stackrel{\smallcirc}{\{\,\cdot\,\}}= d/d\tau$,
the time-dependent oscillator equation \eqref{eqx} is taken into 
\beq
\ccabove(\xi) +\, \bar\omega^2 \xi=0\,,
\where
\bar{\omega}^2=\frac{1}{\dot{\tau}^2}\left(
\omega^2(t)
-\frac{1}{2}\frac{\dddot{\tau}}{\dot{\tau}}
+\frac{3}{4}\left(\frac{\ddot{\tau}}{\dot{\tau}}\right)^2
-\frac{\ddot{\lambda}}{2}-\frac{\dot{\lambda}^2}{4}\right) \,.
\label{eqy}
\eeq 
Thus the problem is reduced to one with \emph{time independent} frequency, $\bar{\omega}$ in \eqref{labjegy} \footnote{ 
Turning off $\lambda$, \eqref{eqy} can be presented, consistently with eqn. (7) of \cite{GWG_Schwarz}, as
\beq
\omega^2(t)={\dot{\tau}^2}\,\bar{\omega}^2 +
\half {\bm S}(\tau)\,
\label{Schwarzform}
\eeq
where 
${\bm S}(\tau)=\frac{\dddot{\tau}}{\dot{\tau}}
-\frac{3}{2}\left(\frac{\ddot{\tau}}{\dot{\tau}}\right)^2
$
is the \emph{Schwarzian derivative} of $\tau$.
} .

Let us now recall the formula \#(19) of \JI in \cite{JunkerInomata} which 
 tells us how  propagators behave under the coordinate transformation $(\xi,\tau) \longleftrightarrow (x,t)$ ~:
\begin{eqnarray}
K_{2}(x'',t''| x',t')=
\left[\left(\frac{\partial \xi'}{\partial x'}\right)\left(\frac{\partial \xi''}{\partial x''}\right)\right]^{\frac{1}{2}}K_{1}(\xi'',\tau''|\xi',\tau')\,.
\label{Inomata19}
\end{eqnarray}
Here  $K_2=K^{ext}$ is the propagator of
an oscillator with time-dependent frequency and friction, $\omega(t)$  and $\lambda(t)$, respectively --- the one we are trying to find.
$K_1$ is  in turn the Maslov-extended propagator of an 
 oscillator with no friction and constant frequency, as in \eqref{constMprop}. 
 Then  the  propagator for the \emph{harmonic oscillator with time-dependent frequency and friction}, eqn. \eqref{JIMaslov},
is obtained using \eqref{Zhao7}.

 Notice that \eqref{JIMaslov} is \emph{regular} at the points $r_k\in J_k$ where $\sin=\pm1$.
 However at caustics,
 $\tau''-\tau' =(\pi/\bar{\omega})\ell$, $K^{ext}$ diverges and we have instead \eqref{JIcaustic}.
 
Henceforth we limit our investigations to $\lambda=0$.

\subsection{Properties of the Niederer map}\label{Npropesec}

More insight is gained from the perspective of the generalized Niederer map \eqref{XTtrans}. We first study their properties in some detail. For simplicity we choose, in the rest of this section, $x'=t'=0$ and $x''\equiv x$ and $t''\equiv t$. 

We start with the observation  that the Niederer map \eqref{XTtrans} becomes singular where the cosine vanishes, i.e., where
\beq
\cos[\baromega\tau(r_k)]=0,
\quad \text{i.e.}\quad 
\tau(r_k)=(k+\half)\frac{\pi}{\bar\omega},
\quad k=0\,,\pm 1,\dots
\label{taurk} 
\eeq 
$r_k<r_{k+1}$ because $\tau(t)$ is an increasing function by \eqref{tauint}. Moreover, each interval  
\beq
I_k= \big[r_k,r_{k+1}\big],
\quad
k= 0, \pm 1, \dots\,
\label{Iinterval}
\eeq
is mapped by \eqref{XTtrans} onto the full range $-\infty < T < \infty$. Therefore the inverse mapping  
 is \emph{multivalued}, labeled by integers $k$,
\beq
N_k~: T \to t=\frac{\arctan_k\baromega{T}}{\baromega}\,,
\quad X \to x = \frac{X}{\sqrt{1+\baromega^2 T^2}}\,,
 \label{Nkdef}
\eeq
where
$ 
\arctan_k (\,\cdot\,) = \arctan_0 (\,\cdot\,) + k\pi
$ 
with $\arctan_0 (\,\cdot\,)$  the principal determination i.e. in $(-\pi/2, \pi/2)$.
\begin{figure} 
\hskip-4mm
\includegraphics[scale=0.4]{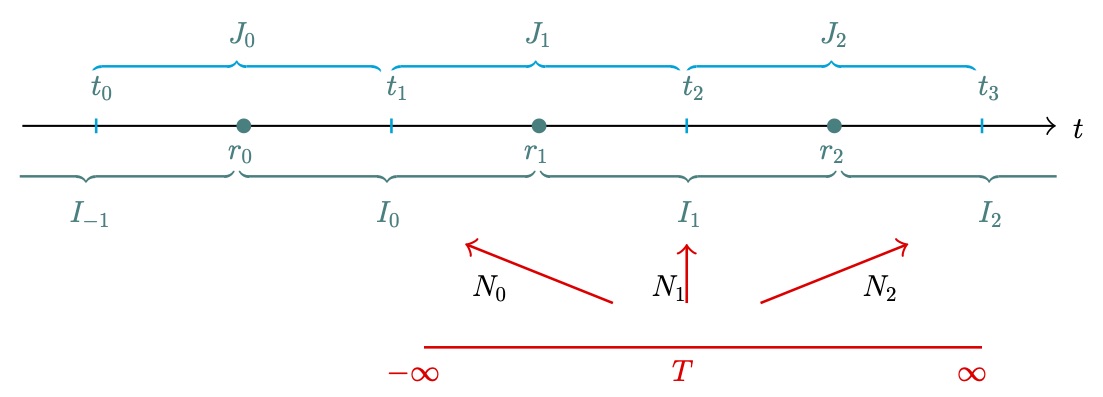}
\vskip-3mm
\caption{\textit{The generelized Niederer map \eqref{XTtrans} maps each interval \dgreen{${\bf I_k=(r_k,r_{k+1})}$} onto the entire real line $-\infty < T < \infty$. Its inverse mapping is therefore multivalued, labeled by an integer $k$.
The classical motions and the propagator
are both regular at the separation points \dgreen{${\bf r_k}$}. All classical trajectories are focused
at the caustic points \cyan{${\bf t_{\ell}}$}, where the propagator diverges.}
\label{IkJlfig}
}
\end{figure}
Then
$\lim_{t\to r_k-}\tan{t}=\infty$ and 
$\lim_{t\to r_k+}\tan{t}=-\infty$ imply that 
\beq
\lim_{T\to\infty}{N_k}(T) =
\; r_{k+1}\;  = \lim_{T\to-\infty}
N_{k+1}(T)\,.
\label{invN}
\eeq
Therefore the intervals $I_k$ and $I_{k+1}$ are joined at $r_{k+1}$ and the $I_k$ form a partition the time axis,
$
\big\{-\infty < t < \infty\big\} ={\cup}_{k} I_k\,.
$
\goodbreak

\smallskip
Returning to \eqref{JIMaslov} (which is  \eqref{constMprop} with $\Omega_0 \Rightarrow \baromega,\, t \Rightarrow \tau$) we then observe that  
whereas the propagator is regular at \dgreen{$r_k$}, it
diverges at caustics,  
\beq
\sin[\baromega\tau(t_{\ell})]=0 \quad \text{i.e.}\quad
\tau(t_{\ell})=\frac{\pi}{\baromega}\,\ell,\quad {\ell}=0,\,\pm1,\dots\, ,
\label{caustic}
\eeq
cf. \eqref{Deltataupiell}.
Thus $t_{\ell} \leq t_{\ell+1}$, and
\beq
N_k(-\infty)=r_k,\quad N_{k}(T=0)= t_{k+1},\quad
N_k(+\infty)=r_{k+1}\,.
\label{Nkrt}
\eeq
Thus $N_k$ maps the full $T$-line into $I_k$ with $t_k$  an internal point. Conversely, $r_k$ is an \emph{internal point} of $J_k$. The intervals $J_\ell= \big[t_\ell,t_{\ell+1}\big]$  cover again the time axis,
$ 
\displaystyle\cup_{\ell} J_{\ell}=\big\{-\infty < t < \infty\big\}\,.
$

By \eqref{xtaueq} the classical trajectories are regular at $t=r_k$. Moreover, for arbitrary initial velocities,
\beq 
\sqrt{\dot\tau(t_{\ell+1})}{\,}x(t_{\ell+1}) =
 - \sqrt{\dot\tau(t_{\ell})}{\,} x(t_{\ell})\,
\label{halfpershift}
\eeq
implying that  after a half-period $\baromega\tau \to \baromega\tau+\pi$ all classical motions are focused  at the same point.
 The two entangled sets of intervals are shown in fig.\ref{IkJlfig}. 
 

The Niederer map  \eqref{NiedEisen} ``E-D lifts'' to Bargmann space.

\kikezd{Theorem 3.} \textit{The E-D lift of the inverse of the Niederer map \eqref{NiedEisen} we shall denote by $\widetilde{N}_k : (X,T,S) \to (x,t,s)$ ($t \in I_k$) is}
\beq
t= \frac{\arctan_k \baromega T}{\baromega}\,, \quad
x = \frac{X}{\sqrt{1+\baromega^2T^2}}\,, \quad
s =  S +  
\frac{X^2}{2}\frac{\bar{\omega}^2T}{1+\bar{\omega}^2T^2}\,. 
\label{xtsXTS}
\eeq
Note that $t$ depends on $k$, $t=N_k(T)$, but $x$ and $s$ do not. 

\vskip2mm
\noindent{\textit{Proof}}~: 
These formul{\ae} follow at once by inverting \eqref{NiedEisen} at once with the cast $\omega_0 \Rightarrow \baromega,\, t \Rightarrow \tau$. 
Alternatively, it could also be proven as for  of Theorem 2.

\smallskip
For each integer $k$ \eqref{xtsXTS} maps the real line $-\infty < T < \infty$ into the  ``open strip'' \cite{BurdetOsci} 
$\big[r_k,r_{k+1}\big]\times\IR^2 \equiv I_k\times\IR^2$
with
 $r_k$ defined in \eqref{taurk}. Their union
covers the entire Bargmann manifold of the oscillator.

Now we pull back the free dynamics by the multivalued inverse \eqref{xtsXTS}. We put $\baromega=1$ for simplicity.
The free motion with initial condition $X(0)=0$,
\beq
X(T) = a T, \quad 
S(T) = S_0 -\frac{a^2}{2}T\,,
\label{freelift}
\eeq
E-D lifts by \eqref{xtsXTS} to
\beq
x(t) = a \sin t
\qquad
s(t) = S_0 - \frac{a^2}{4}\sin 2t\,,
\label{pullbacks}
\eeq
consistently with $s(t)=s_0-\bar{\cal A}_{osc}$, as it can be checked directly.
Note that the $s$ coordinate oscillates with doubled frequency. 
\begin{itemize}
\item
At  $t= r_k=(\half+k)\pi$ (where the Niederer maps are joined), 
we have,
$ 
\lim_{t\to r_{k}} x(t) = (-1)^{k+1}a\,,
\;
\lim_{t\to r_{k}} s(t) = S_0.
$ 
Thus the pull-backs of the Bargmann-lifts of free motions  are glued to smooth curves.

\item 
Similarly at t caustics $t=t_\ell=\pi\ell$ we infer from \eqref{pullbacks} that for all initial velocity $a$ and for all 
$\ell$
$ 
\lim_{t\to t_{\ell}}
x(t) = 0, 
 \,
\lim_{t\to t_{\ell}}s(t) = S_0\,. 
$ 
Thus the lifts are again smooth at $t_\ell$ and after each half-period all motions are focused above the initial position $(x(0)=0, s(0)=S_0)$.

\end{itemize}

\subsection{The propagator by the Niederer map}\label{Npropsec}

Now we turn at the quantum dynamics.
Our starting point is the free propagator \eqref{freeprop} which (as mentioned before) is valid only for 
$0< T''-T'$. Its extension to all $T$ involves the \emph{sign} of $(T''-T')$ \cite{BurdetOsci}. 

Let us explain this subtle point in some detail. First of all, we notice that the usual expression \eqref{freeprop} involves a square root which is double-valued, obliging us to \emph{choose} one of its branches. Which one do we choose is irrelevant -- it is a mere gauge choice. However once we do choose one, we must stick to our choice. Take, for example, the one for which $\sqrt{-i}=e^{-i\pi/4}$ --- then the prefactor in \eqref{freeprop} is 
$$
\left[\frac{1}{2\pi{i}\hbar(T''-T')}\right]^{1/2}=
e^{-i\pi/4}\left[\frac{1}{2\pi\hbar \big|T''-T'\big|}\right]^{1/2}.
$$
Let us now consider what happens when $T''-T'$ changes sign. Then 
the prefactor gets multiplied by $\sqrt{-1}\,$ so it becomes, \emph{for the same choice of the square root}, 
\beq
e^{i\pi/2\,}e^{-i\pi/4}\left[\frac{1}{2\pi\hbar \big|T''-T'\big|}\right]^{1/2}
=e^{+i\pi/4}\left[\frac{1}{2\pi\hbar \big|T''-T'\big|}\right]^{1/2}.
\label{nulljump}
\eeq
In conclusion, the formula valid for all $T$ is,
\beqa
K_{free}(X'',T''|X',T') =e^{-{i}\smallover{\pi}/{4}\,{\rm sign}(T''-T')}
\left[\displaystyle\frac{1}{2\pi\hbar |T''-T'|}\right]^{1/2}
\!\!
\exp\left\{\displaystyle\frac{i}{\hbar}\bar{\cal A}_{free}\right\}\,,
\label{truefreeprop}
\eeqa
 where
\begin{equation}
\bar{\cal A}_{free}=\frac{(X''-X')^2}{2(T''-T')}\,
\end{equation}%
is the free action calculated along the classical trajectory.
Let us underline that \eqref{truefreeprop} already involves a ``Maslov jump'' $e^{-i\pi/2\,}$ -- which, for a free particle, happens at $T=0$.
For $T''-T'=0$ we have $K_{free}=\delta(X''-X')$.

Accordingly, the wave function $\Psi\equiv \Psi_{free}$ of a free particle is, by \eqref{wfprop}, 
\begin{eqnarray}
\Psi\left(X'',T''\right)\!=\!
e^{-i\frac{\pi}{4}\!\sign\left(T''-T'\right)}
\left[\frac{1}{2\pi \hbar |T''-T'|}\right]^{1/2} 
\!\!
\int_{\IR}\exp\left\{\frac{i}{\hbar}\bar{\cal A}_{free}\right\}\Psi \left(X',T'\right) dX'\,.\quad
\label{freewf}
\end{eqnarray}%


Now we pull back the free dynamics using the multivalued inverse Niederer map.     
It is sufficient to consider the constant-frequency case $\baromega=\const$ and denote time by $t$. 
Let $t$ belong to the range of $N_k$ in \eqref{Nkdef}, 
$
t\in I_k=[r_k,r_{k+1}] = N_k\big(\{-\infty < T < \infty\}\big).
$
Then applying the general formulae in sec.\ref{BJIsec}  yields \cite{BurdetOsci}, 
\beq
\begin{array}{lll}
\wpsi(x'',t'',s'')&=&\cos^{-1/2}[\baromega(t''-t')]
\widetilde{\Psi}(X'',T'',S'') = e^{-\frac{i\pi}{4}\sign\left(\frac{\tan\baromega(t''-t'')}{\baromega}\right)}\times
\\[8pt] 
&&\cos^{-1/2}\left[\baromega(t''-t')\right]
\exp\left(\frac{i}{\hbar}s''\right)\exp\left(-\frac{i}{\hbar}(\frac{1}{2}\baromega {x''}^{2}\tan[\baromega(t''-t'')]\right)
\\[8pt]
&&\quad\sqrt{\displaystyle\frac{|\baromega|}{2\pi\hbar
|\tan[\baromega(t''-t')]|}}%
\displaystyle\int_{\IR}\!\exp\left\{\frac{i}{\hbar}\frac{{\baromega}|\frac{{x''}}{\cos[\baromega
(t''-t')]}-x'|^{2}}{2\tan[\baromega(t''-t')]}\right\}
\psi(x',t') dx'\,.
\end{array}%
\nn
\eeq
However the second exponential in the middle line
 combines with the integrand in the braces  in the last line
to yield \emph{the  action calculated along the classical  oscillator trajectory},
\beq
\bar{\cal A}_{osc}=  \frac{\baromega}{2\sin\baromega(t''-t')}%
\big(({x''}^{2}+{x'}^{2})\cos\baromega(t''-t') -2x''x'\big)\,. 
\eeq
Thus using the equivariance we end up with,
\begin{eqnarray}
\psi_{osc}\left(x'',t''\right) &=&\cos^{-1/2}[\baromega(t''-t')]\, \exp\left[
-\frac{i\pi}{4}\sign\left(\frac{\tan[\baromega(t''-t')]}{\baromega}\right) \right] \times \quad
\label{wffromfree} 
\\[6pt]
&&\sqrt{\frac{|\baromega|}{2\pi\hbar |{\tan[\baromega(t''-t')]}|}} 
\int_{\IR}\exp\left\{\frac{i}{\hbar}\bar{\cal A}_{osc}
\right\} \psi_{osc}\left(x',t'\right) dx\,.
\nonumber
\end{eqnarray}  
Now we recover the Maslov jump which comes from the first line here. For simplicity we consider again $t'=0,\,x'=0$ and denote $t''=t,\, x''=x$.

Firstly, we observe that the conformal factor $\cos\baromega t$  has constant sign in the domain  $I_k$ and changes sign at the end points. In fact, 
\beq
\cos \baromega t =\left(-1\right)^{k+1}|\cos \baromega t|
\Rarrow
\cos^{-1/2}\left(\baromega t \right) =e^{-i\smallover{\pi}/{2}(k+1)} |\cos \baromega t |^{-1/2}.
\label{cosk}
\eeq%
The cosine enters into the van Vleck factor while
the phase  combines with $\exp\left[-%
\frac{i\pi}{4}\,\sign(\frac{\tan\baromega t}{\baromega}) \right]$.
Recall now that  
$
t_{k+1} = N_k\big({T=0}\big)\,$
divides $I_k$ into two pieces, 
$
I_k=[r_k, t_{k+1}] \cup [t_{k+1}, r_{k+1}],
$
cf. fig.\ref{IkJlfig}.
But $t_{k+1}$ is precisely where the tangent changes sign~: this term contributes to the phase  in $[r_k, t_{k+1}]$ $-\pi/4$, and $+\pi/4$ in $[t_{k+1}, r_{k+1}]$.
Combining the two shifts, we end up with the phase
\beq\bigbox{
\begin{array}{cll}
 -\frac{\pi}{4}\big(1+ 2\ell\big)  \quad &\text{\small for} \quad &\;r_k \;\;  <  t < t_{k+1}
\\[6pt]
-\frac{\pi}{4}\big(1+ 2(\ell+1)\big) 
\qquad &\text{\small for}\quad &t_{k+1} < t  <  r_{k+1}
\end{array}
\;\;\where
\ell =  {\rm Ent}\left[\frac{\baromega\tau}{\pi}\right] =k+1 \,
}
\label{Maslovjump}
\eeq
which is the Maslov jump at $t_{\ell}$.

Intuitively, that the multivalued $N_k$ ``exports'' to the oscillator at $t_{\ell+1}$ the  phase jump of the free propagator at $T=0$. Crossing from $J_\ell$ to $J_{\ell+1}$
shifts the index $\ell$ by one.

\section{Probability density and phase of the propagator: a picturial view}\label{propsec}

\subsection{For constant frequency}\label{Cfreqpic}

We assume first that the frequency is constant.  
We split the propagator $K(x,t)\equiv K(x,t|0,0)$ in \eqref{constMprop} as, 
\beq
K(x,t)=|K(x,t)|\,P(t),\qquad 
P(t)=e^{i(phase)}.
\label{Ksplitting}
\eeq
The \emph{probability density},
\beq
|K(x,t)|^2=
\frac{\Omega_0}{2\pi \hbar\,\big|\sin \Omega_0t\big|}
\eeq
viewed as a surface above the $x - t$ plane,
 diverges at $t=t_{\ell}=\pi \ell$, $\ell=0,\pm1,\dots$.

\begin{figure} 
\hskip-5mm
\includegraphics[scale=0.49]{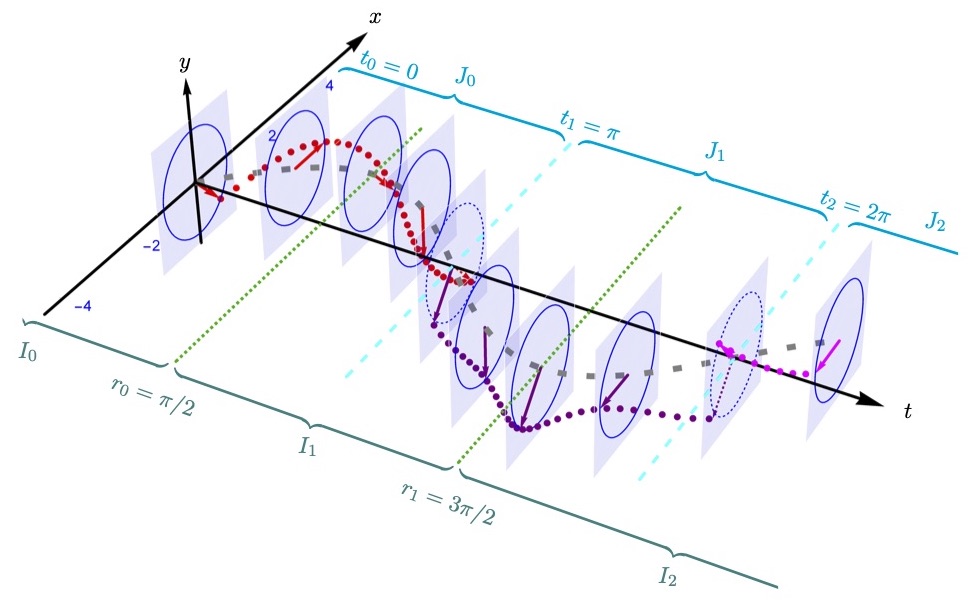}
\vskip-4mm
\caption{\textit{The phase factor ${\bf P}(t)$  of the propagator in \eqref{Ksplitting} lies on the unit circle of the complex plane plotted vertically along a classical path 
$\bar{\gamma}(t)$. 
The orentation is  positive if it is clockwise when seen from $t = +\infty$.
In the time interval $J_{\ell}$ labeled by the Maslov index $\ell={\rm Ent}[t/\pi]$, the factor ${\bf P}(t)$ precesses around ${\bf P}_\ell=\exp[-i\frac{\pi}{4}(1+2\ell)]$ with double frequency w.r.t. the classical path, $\bar{\gamma}(t)$.
Arriving at a caustic the phase jumps by $(-\pi/2)$ (\red{\bf red} becoming \purple{\bf purple}) and then continues until the next caustic when it jumps again (and becomes \magenta{\bf magenta}), and so on.} 
\label{Maslovphase}
}
\end{figure}

Representing the \emph{phase of the propagator} would require $4$ dimensions, though. However, recall that that the dominant contribution to the path integral should come from where the phase is stationary \cite{FeynmanHibbs}, i.e., from the neighborhood of classical paths $\bar{x}(t)$, distinguished by the vanishing of the first variation, $\delta{\cA}_{\bar{x}}=0$.
 Therefore we shall study the evolution of the phase along classical paths  $\bar{x}(t)$ for which
 \eqref{xtaueq} yields,
for $\hbar=\bar{\omega}=1$ and $a\in\IR,\, b=0$,
\beq
\bar{x}_a(t)= a\sin t
\aand 
P_a(t) = \exp \left\{-\frac{i\pi}{4}\big[1-\frac{a^2}{\pi}\sin 2t\big] -\frac{i\pi}{2}\ell\right\}\,,
\label{xaP}
\eeq
depicted in Fig.\ref{Maslovphase}. 
\goodbreak
\goodbreak

An intuitive understanding  comes by noting 
 that when $t \neq \pi \ell=t_\ell$, then different initial velocities $a$ yield classical paths $\bar{x}_{a}(t)$s with different end points, and thus contribute  to different propagators.
However approaching from the left $\ell$-times a half period,  
$
t\to (\pi\,\ell)-\,,
$ 
all classical paths get  focused at the same end-point ($x=0$ for our choice) and for all $a$,
\beq
P_{a}(t\to \pi\ell-) = e^{-i\frac{\pi}{4}(1+{2}\ell)}
\equiv  P_{\ell}
\,.
\label{llimitP}
\eeq
which is precisely the Maslov phase.
Thus  \emph{all classical paths contribute equally}, by $P_{\ell}$, and to the \emph{same} propagator. Comparing with the right-limit, 
\beq
P_{a}(t\to \pi\ell+) = e^{-i\frac{\pi}{4}(1+{2}(\ell+1)}
 = 
 P_{\ell+1} = e^{-\frac{i\pi}{2}}P_{\ell}.
\label{rlimitP}
\eeq
the Maslov jump is recovered.
Choosing instead $y\neq 0$ there will be no classical path from $(0,0)$ to $(y,\pi\ell)$ and thus no contribution to the path integral.
\goodbreak

To conclude this section we just mention with that the extended Feynman method \cite{HFeynman} with the cast 
$\baromega$ = constant frequency and $\tau$ = ``fake time''  would lead also to \eqref{JIMaslov} and \eqref{JIcaustic} with the integer $\ell$ counting the number of negative eigenvalues (Morse index) of the Hessian \cite{Schulman,semiclassic,Maslov}.  

\goodbreak

\subsection{A time-dependent example: the Mathieu equation}\label{Mathieusec}

\begin{figure} [ht]
\includegraphics[scale=0.265]{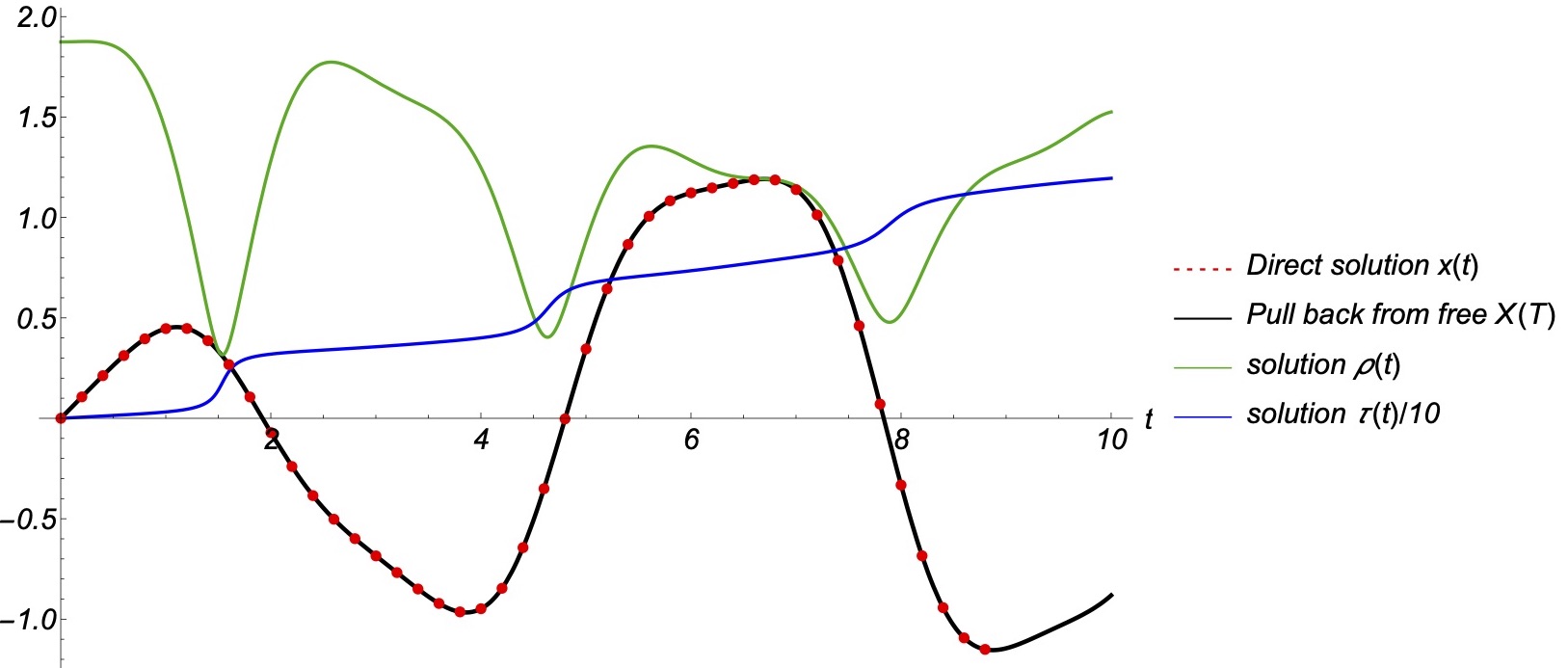}
\vskip-2mm
\caption{\textit{The analytic solution of the Mathieu equation with  $a=2,q=1$ for $x(t)$ (\red{\bf dotted} in \red{\bf red}), lies on the {\bf black} curve got by  \eqref{xAnsatz} from combining the numerically obtained \dgreen{$\rho(t)$} (in \dgreen{\bf green}) and \blue{$\tau(t)$} (in \blue{\bf blue}), which are solutions of the pair  \eqref{genErmakov}-\eqref{tauint}. 
The  {\bf black} curve is also obtained
 by pulling back the free solution \eqref{aTb} by the inverse Niederer map \eqref{Nkdef}.}
\label{xrho1}
}
\end{figure}

The combined Junker-Inomata - Arnold method allows us to go  beyond the constant-frequency case, as illustrated here for no friction or driving force, $\lambda=F \equiv0$, but with explicitly time-dependent frequency. For $\Omega^2(t) = a-2q\cos 2t$, for example, \eqref{SLeq} becomes the Mathieu equation,
\beq
\ddot{x}+(a-2q\cos 2t)x=0\,.
\label{Mathieueq}
\end{equation}
This equation can be solved either 
analytically using Mathieu functions \cite{Mathieuf},
or numerically, providing us for $a=2$ and $q=1$ (for which odd Mathieu functions are real) with the dotted curve (in \red{\bf red}), shown in Fig.\ref{xrho1}. 

Alternatively, we can  use the Junker-Inomata -- Arnold transformation \eqref{ArnoldXT} \cite{LopezRuiz,Arnold}. 
We first  achieve 
 $\bar{\omega}=1$ by a redefinition, $\tau\to
\tau^{\prime}=\bar{\omega}\tau $.
Inserting the Ansatz \eqref{xAnsatz} into \eqref{Mathieueq}  yields the pair of coupled equations
\eqref{eqrho}-\eqref{trlconstr}.
We choose $u_{p}=0$ and two independent solutions $u_{1}(t)$ and $u_{2}(t)$, 
\eqref{u1u2rt}, with initial conditions \eqref{initcond} with $t_0=0$ i.e.,
$ 
\tau(0)= \dot{\rho}(0) =0,\, \rho(0) =\dot{\tau}(0) =1\,,
$ 
which fix the integration constant, 
$
C
=\rho^{2}\left({0}\right)\dot{\tau}({0})=1.
$ 
 Then, consistently with the general theory outlined above, the Arnold map \eqref{ArnoldXT} lifted to Bargmann space becomes  \eqref{XTtrans}, completed with \eqref{SJI} with $\lambda=0$. %
  
Eqn \eqref{rhotaueq} is solved by following the strategy outlined in sec.\ref{JIsec}. 
Carrying out those steps numerically
provides us with Fig.\ref{xrho1}.

\begin{figure} [ht]
\includegraphics[scale=0.37]{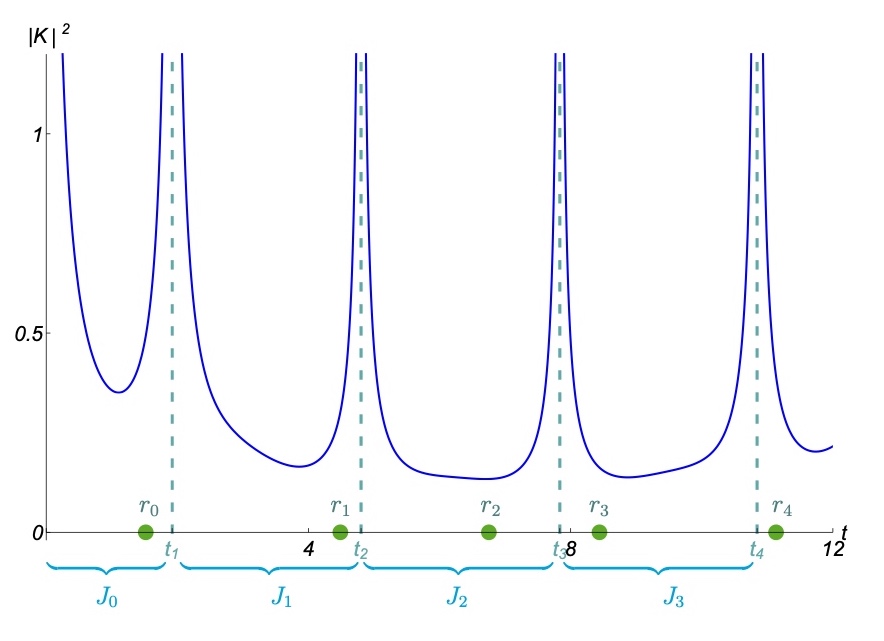}
\vskip-6mm
\caption{\textit{The probability density $|K(x,t)|^2$  \eqref{probadens} does not depend on $x$ and is regular in each interval $\cyan{{\bf J}_{\ell}}$ between the  adjacent points $\cyan{\bf t_{\ell}}$ \eqref{caustic}, where it diverges.
The $\dgreen{\bf r_k}$ which determine the domains $\dgreen{{\bf I}_k}$ of the generalized Niederer map \eqref{XTtrans} lie between the $\cyan{\bf t_{\ell}}$ and conversely.
} 
\label{MathieuK2}
}
\end{figure}
\begin{figure}
\includegraphics[scale=0.474]{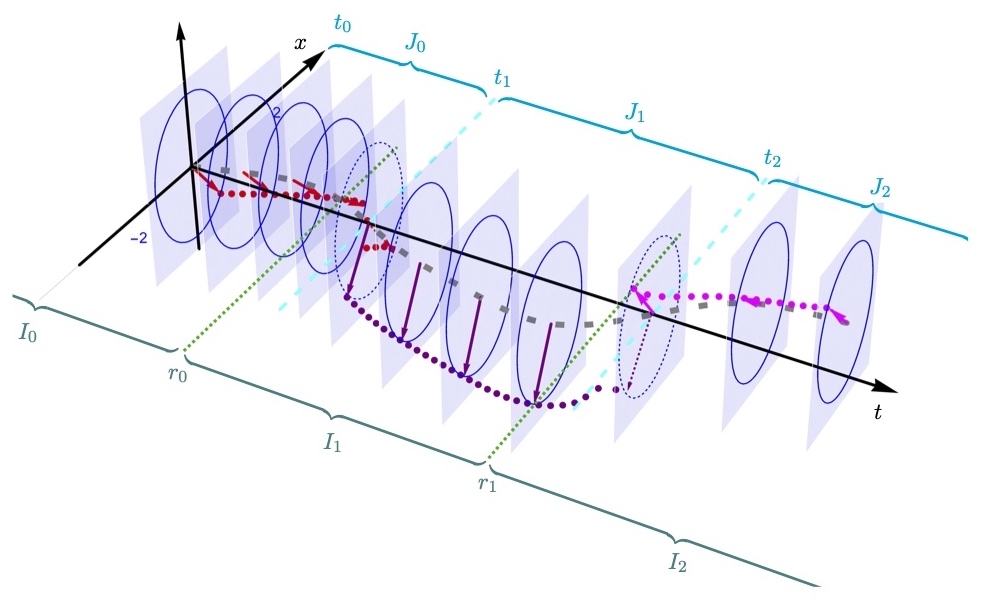}
\vskip-4mm
\caption{\textit{For $0 < t < t_1$ the Mathieu phase factor ${\bf P}(t)$ plotted along a classical path $\bar\gamma(t)=(\bar{x}(t),t)$ precesses around  $e^{-i\pi/4}$.
Arriving at the caustic point \cyan{$\tau(t_1)=\pi$} its phase jumps by $(-\pi/2)$, then oscillates around $e^{-3i\pi/4}$ until \cyan{$\tau(t_2)=2\pi$}, then jumps again, and so on. 
}
\label{MathieuPhase}
}
\end{figure}

From the general formula \eqref{JIprop} we deduce, for our choice $x''=x,t''=t, x'=t'=0$, the probability density \footnote{The wave function is multiplied by the square root of the conformal factor, cf. \eqref{otoconst}.} 
\beq
|K(x,t)|^2=\frac{\sqrt{\dot{\tau}}}{2\pi\hbar |\sin\tau(t)|}\,,
\label{probadens}
\eeq
 happens not depend on the position, and can therefore be plotted  as in Fig.\ref{MathieuK2}.
The propagator $K$ and hence the probability density \eqref{probadens} diverge at $\cyan{\bf t_{\ell}}$, which are roughly
$
t_1 \approx 1.92,\,
t_2 \approx 4.80,\,
t_3 \approx 7.83\,.
$
The classical motions are regular at the caustics,
$ 
\bar{x}(t_{\ell}) \propto  
 \rho(t_{\ell}) \approx 0,
$ 
see sec. \ref{Maslovsec}.
The domains \cyan{${\bf I_k} = [r_{k-1},r_k]$} of the inverse Niederer map are shown in fig.\ref{MathieuK2}.
 Approximately,
$
r_1  \approx  1.52,\, 
r_2  \approx  4.49,\,
r_3  \approx 6.75,\,
r_4  \approx 8.44\,.
$
The evolution of the phase factor along the classical path is depicted in fig.\ref{MathieuPhase}.

\goodbreak
\section{Conclusion}

The Junker-Inomata -- Arnold approach  yields (in principle) the exact propagator for any quadratic system by switching from \emph{time-dependent} to \emph{constant frequency} and redefined time, 
\beq
\omega(t) \;\to\; \baromega=\const
\aand t \;\to\; \text{``fake time''}\;\; \tau\,.
\eeq 
The propagator \eqref{JIMaslov}-\eqref{JIcaustic} is then derived from the result known for constant frequency. A straightforward consequence is the Maslov jump for arbitrary time-dependent frequency $\omega(t)$~: everything depends only on the product $\baromega\,\tau$.

By switching from $t$ to $\tau$ the Sturm-Liouville-type
difficulty is not eliminated, though, only transferred to that of finding $\tau=\tau(t)$ following the prcedure outlined in sec.\ref{JIsec}. We have to solve first solve EMP equation \eqref{genErmakov}  for $\rho(t)$ (which is non-linear and has time-dependent coefficients),  and then integrate $\rho^{-2}$,  see  \eqref{tauint}. Although this is as difficult to solve as solving the Sturm-Liouville equation, however it provides us with theoretical insights. 

When no analytic solution is available, we can resort to numerical calculations. 

\goodbreak
The Junker-Inomata approach of sec.\ref{JIsec}
is interpreted as a Bargmann-conformal transformation between time-dependent and constant frequency metrics, see eqn \eqref{otoconst}.

Alternatively, the damped oscillator can be converted to a free system  by the generalized Niederer map \eqref{XTtrans}, whose  Eisenhart-Duval lift \eqref{ArnoldXT}-\eqref{Sfroms} carries the {conformally flat} oscillator metric  \eqref{Arnoldconfflat} to flat Minkowski space.

Two sets of points play a distinguished r\^ole in our investigations~:
the $r_k$ in  \eqref{taurk}  and the $t_{\ell}$ in
\eqref{caustic}. The $r_k$ divide the time axis into  domains $I_k$ of the (generalized) Niederer map \eqref{XTtrans}. Both classical motions and quantum propagators are \emph{regular} at $r_k$ where these intervals are joined.
The $t_{\ell}$ are in turn  the caustic points where all  \emph{classical trajectories are focused} and the \emph{quantum} propagator becomes \emph{singular}. 

While the  ``Maslov phase jump'' at caustics is well established when the frequency is constant, $\omega=\omega_0=\const$, its extension to the time-dependent case $\omega=\omega(t)$ is more subtle. In fact, the proofs we are aware of 
  \cite{Arnold67,SMaslov,Burdet78,RezendeJMP} use sophisticated mathematics, or a lengthy direct calculation of the propgagator \cite{Cheng87}. A bonus from the Junker-Inomata transcription \eqref{Inomatatransf} we follow here is to provide us with a straightforward extension valid to an arbitrary 
$\omega(t)$. Caustics arise when \eqref{Deltataupiell} holds, and then the phase jump is given by \eqref{Maslovjump}. 

The subtle point mentioned above comes from
the standard (but somewhat sloppy) expression \eqref{freeprop} which requires to choose a  branch of the double-valued square root function. Once this is done, the sign change of $T''-T'$ induces  a phase jump $\pi/2$. 
Our  ``innocently-looking'' factor \emph{is} in fact the  Maslov jump for a free particle at $T=0$  (obscured when one considers the propagator  for $T>0$ only). Moreover, it then becomes the key tool for the ocillator~: intuitively, the multivalued inverse Niederer map repeates, all over again and again, the same jump. 
Details are  discussed in sec.\ref{Maslovsec}. 
 
The transformation \eqref{Inomatatransf} is related to the  \emph{non-relativistic ``Schr\"odinger'' conformal symmetries} of a free non-relativistic  particle \cite{Jackiw72,Niederer72,Hagen72} 
later extended to the oscillator \cite{Niederer73} and an inverse-square potential \cite{Fubini}. These results can in fact be derived using a time-dependent conformal transformation of the type \eqref{Inomatatransf} \cite{BurdetOsci,GWG_Schwarz}.

The above results are readily generalized to higher dimensions. For example, the oscillator frequency can be time-dependent,  uniform electric  and magnetic fields and a curl-free ``Aharonov-Bohm'' potential (a vortex line \cite{Jackiw90}) can also be added \cite{DHP2}. 
Further generalization involves a Dirac monopole  \cite{Jackiw80}.

Alternative ways to relate free and harmonically trapped motions are studied, e.g., in \cite{Andr18,Inzunza,Dhasmana}. Motions with Mathieu profile are considered also in \cite{Guha21}.

%

\vspace{-5mm} 
\begin{acknowledgments}\vskip-4mm
We are indebted to Gary Gibbons and to Larry Schulman for correspondence and advice.
This work was partially supported by  the National Natural Science Foundation of China (Grant No. 11975320). 
\end{acknowledgments}

\goodbreak

\end{document}